# Assessing Mobile Learning System Performance in Indonesia: Reports of the Model Development and Its Instrument Testing

Aang Subiyakto[1, a)], Noni Erlina[1, b)], Yuni Sugiarti[1, c)], Nashrul Hakiem[1, d)], Moh. Irfan[2, e)], and Abd. Rahman Ahlan[3, f)]

[1]*UIN Syarif Hidayatullah Jakarta Jl. Juanda 95, Kota Tangerang Selatan, 15412, Indonesia*
[2]*UIN Sunan Gunung Djati Bandung Jl. A.H. Nasution 105, Bandung, 40614, Indonesia*
[3]*International Islamic University Malaysia Islamic University Jl. Gombak, Kuala Lumpur, 53100, Malaysia*

[a)]Corresponding author: aang_subiyakto@uinjkt.ac.id
[b)]noni.erlina16@mhs.uinjkt.ac.id
[c)]yuni.sugiarti@uinjkt.ac.id
[d)]hakiem@uinjkt.ac.id
[e)]irfan.bahaf@uinsgd.ac.id
[f)]arahman@iium.edu.my

**Abstract.** It is undeniable that people life patterns and technological developments are interrelated within a supply and demand cycle. In the education world, the emergence of the internet and mobile technologies has opened the learning boundaries through the use of mobile learning (m-learning). In Indonesia, the learning service industry has been begun to enliven the outside school education sector for almost five years ago. Even though the learning has been discussed around a decade ago, however, it is still rare studies that discuss the performance of the m-learning system based on the end-user perceptions in particular. Therefore, the study may still indispensable, especially from the perspectives of a developing nation. This paper elucidates the preliminary stage results of the above-mentioned study, including the results of the model development and its instrument testing. The DeLone and Mclean's information system (IS) success model was adopted, combined with the individual motivation and organizational culture theories, and then adapted into the processional and causal logic of the success model. Around 50 respondent data were collected online and processed and analyzed based on the outer model assessments of the PLS-SEM method using SmartPLS 3.0 to know the reliability and validity of each indicator. The result shows that two of 31 are rejected indicators. The rejections may be the revision considerations for the next study stages. Although this may be trivial for experts, the clarity of its methodological explanations may guide the novice researchers, how to develop a research model and its instrument testing.

## INTRODUCTION

It is undeniable that developments of internet and mobile phone technologies have changed many fields in daily human life, including in the education world [1-3] . One of the forms are electronic-based learning system, ranging from electronic learning (e-learning), ubiquitous learning (u-learning), and mobile learning (m-learning). In the last decade, several mobile learning service providers have been and have started to be used and have their own market share in Indonesia [4]. However, it is a tendency that assessment of the above-mentioned learning systems still tends to be limited. In addition, the assessments are still carried out based on the technological development perspective or conducted in the context of social point of view of the psychological and social side. The social computing investigation that combines the two perspectives above is still rarely seen. On the other hand, although most of the full research articles on the social computing fields also explain the model and its hypothesis developments, it is still rare study which elucidate clearly the conceptual framework of model development, its break down step into the instruments until how to test the instrument. For this reason, it is indispensable that the above-mentioned study was carried out by the authors. The aims were to develop a success model of mobile learning systems and find out the reliability and validity of the instrument. Two research questions were proposed here for guiding the research implementation.

Q1: How to develop m-learning system success model based on IS success models?
Q2: Is the research instrument broken down from the model reliable and valid?

This article reports stages of the model development and its instrument within four parts. Continuing the introductory part of this paper, the second part briefly presents the methodological points of the study implementation. This paper then concludes with the conclusion part at the end of the paper.

## METHOD

This study was conducted within its four main phases, including: the preliminary studies, model development, its operationalization into the measurement items, and the questionnaire development and testing phases. Fig. 1 shows the sequential procedure of the above-mentioned phases.

First, the researchers studied a number of previous studies related to the research programs, selected appropriate theories or models, and then constructed the related theories and models into a conceptual framework. The framework was developed in the context for presenting the interrelated ideas among the theoretical bases used in this study. Fig. 2(a) demonstrates the interrelationship of the conceptual framework.

Second, based on the developed conceptual framework in the prior phase, the model was then developed through adoption, combination, and adaptation of the selected theories and models in the context of the phenomenon which had been the research focus. The rationale of this phase is indications of a number of previous model development assumptions [5-7]. Fig. 2(b) shows the developed model based on its development assumptions.

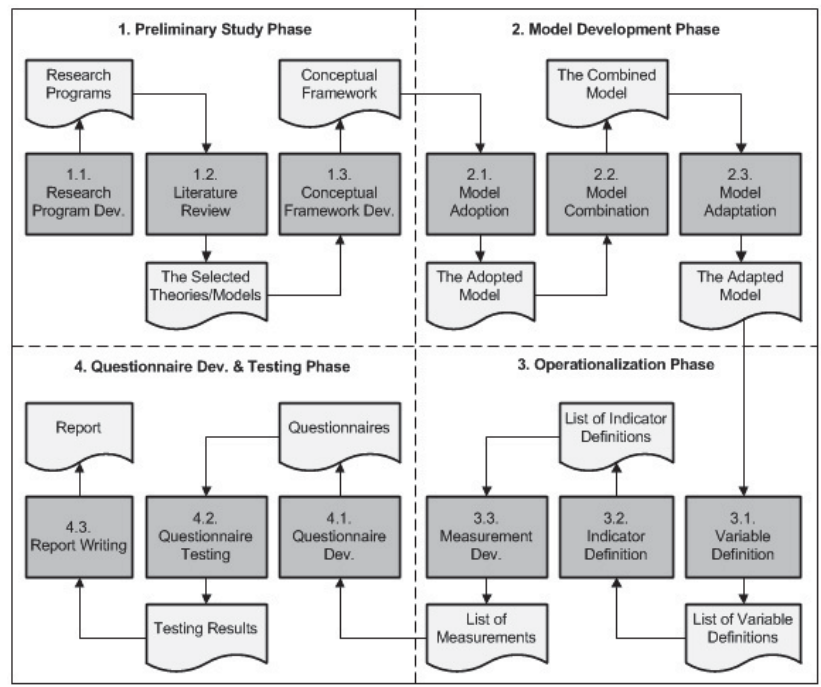

**FIGURE 1.** Research procedure

Third, following to the model development phase, the developed model was then broken down into operational level by defining the variables and their indicators and creating the measurement of each indicator [8] . Table 1 shows list of the 31 indicators and each measure.

Fourth, the measures were then being the main source of the questionnaire development with almost 10 profile questions. The questionnaires were then distributed using online survey with Google Form via social media (Facebook, Whatsapp, Istagram, Twitter, Telegram, and Line). The people were selected based on the snowball purposive sampling. About 50 valid data of the responded respondents were used in the data analysis phase. The researchers used PLS-SEM analysis method with SmartPLS 2.0 in the analysis phase, in regard to the sample size and its powerful analysis points [9-11].

In short, besides the above-mentioned descriptions may present transparently the systematic, cohesive, coherent, comprehensive points of the research implementation; the elucidation may also express reliably the meta-inferences points of the results. Both transparency and reliability points were designed in order to guarantee the research quality, as it was indicated by Eddy, Hollingworth [12] and Subiyakto, Ahlan [13].

## RESULT AND DISCUSSION

### Conceptual Framework

The development of internet and mobile phone technologies has become a catalyst for human life change in various fields, including in the education world [1-3]. Learning that initially used the physical face-to-face model has combined or switched to the electronic ones by implementing the electronic, mobile, and ubiquitous learning concepts. By using IT, the learning stakeholders can be accessed easily as if information is only at the tip of a finger.

On the other hand, it is important to know whether mobile learning system has a positive impact as expected. In this study the researchers adopted the IS success model [14, 15], combined with the two learning aspects (i.e., learning contents and motivation of the learning participant) [2, 16, 17], and then adapted it in terms of mobile learning context [18-22]. The researchers assumed, despite the success model is the popular model for assessing IS performance among in the IS research field; of course, the model may also need to be extended referring to the research context. Therefore, framework of adoption, combination, and adaptation may also have needed here. Fig. 2a presents the conceptual framework used for developing the mobile learning success model.

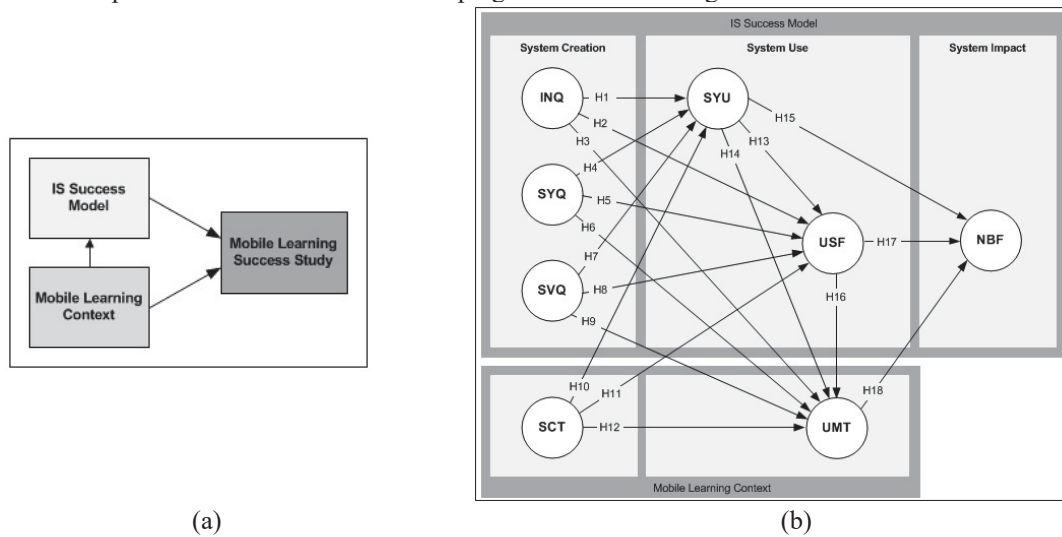

(a)  (b)

**FIGURE 2.** (a) Conceptual framework of the model development, (b) The proposed research model in this study

In the short discussion, we can see that the conceptual framework was developed by considering recommendations of the IS success model themselves [14]. The extension possibility may be performed in terms of the new model development [5]. Moreover, Zwikael and Ahn [23] and Martinsuo [22] indicated that the contextual issues are also part of the essential factors in many technology implementation projects. Therefore, adoption of the mobile learning context may have reasonable here.

### Mobile Learning Success Model and Its Instruments

The developed model was consisted of eight variables. The six ones were adopted from the DeLone and Mclean's IS success model [14,24,25], i.e., information quality (INQ), system quality (SYQ), service quality (SVQ), system use (SYU), user satisfaction (USF), and net benefits (NBF). The two rest variables were adopted in terms of the mobile learning contexts, i.e., system contents (SCT) [19, 20] and user motivation (UMT) [2]. Fig. 5b demonstrates the developed model. The placement of each variable was appropriated to the conceptual framework of the model development.

Figure 2b demonstrated lists the proposed hypotheses in this study. Table X shows list of the indicators and each of the measure. The following descriptions below describe the hypothesis development.

First, in the context of learning system studies in Indonesia [15,26-30] and the similar research among the developing countries [17,31-35], the researchers then proposed nine hypotheses (i.e., H1, H2, H4. H5, H7, H8, H13, H15, and H17) among variables of the IS Success model adopted in this study [14,24,25].

Second, in regard to the adoption of mobile learning context [2,19,20,36,37]; the scholars proposed nine hypotheses, i.e., H3, H6, H9, H10, H11, H12, H14, H16, and H18. These hypotheses were developed for assessing relationships among the contextual variables of mobile learning with the IS success variables.

**TABLE 1.** List of indicators and its measures

| Codes | Names | Measures |
|---|---|---|
| INQ1 | *Timely* | The information displayed is up to date |
| INQ2 | *Usefulness* | The information displayed is useful |
| INQ3 | *Completeness* | The information displayed is complete |
| INQ4 | *Relevancy* | The information displayed is relevant |
| INQ5 | *Accuracy* | The information displayed is accurate and precise |
| SYQ1 | *User friendly* | The system has a display that is easy to operate |
| SYQ2 | *Ease of accessibility* | The system does not take long to access |
| SYQ3 | *Ease of learning* | The system is easy to learn |
| SYQ4 | *Ease of use* | The system has features that are easy to use |
| SYQ5 | *Reliability* | The system rarely experiences errors |
| SVQ1 | *Usage guide* | The system provides usage guidelines |
| SVQ2 | *Responsiveness* | The system gives a quick response when I need help |
| SVQ3 | *Accessibility* | The system can be accessed anywhere and anytime |
| SCT1 | *appropriateness* | The system provides content that is desirable and needed |
| SCT2 | *Timeliness* | The system provides the latest content |
| SCT3 | *Sufficiency* | The system provides quite diverse content |
| SYU1 | *Purpose of use* | The system fits the purpose that I want |
| SYU2 | *Level of use* | The system according to the level of ability that I have |
| SYU3 | *Recurring use* | I often use the system repeatedly |
| SYU4 | *Expectation/Belief* | The system is in line with my expectations |
| USF1 | *Perceived Usefulness* | I feel the benefits of the existence of the system |
| USF2 | *Overall satisfaction* | I feel satisfied with the existence of the system |
| USF3 | *Enjoyment* | I feel comfortable using the system |
| USF4 | *Display interface* | I am interested in using the system because it looks interesting |
| UMT1 | *Expectation* | The system is in accordance with the expectations that I want |
| UMT2 | *Instrumentalist* | The system gives success to the tasks that I have |
| UMT3 | *Valence* | The system delivered results that exceeded my expectations |
| NBF1 | *Efficient* | The system makes working on tasks faster |
| NBF2 | *Effective* | The system makes my job better |
| NBF3 | *Problem solution* | The system helps reduce task errors |
| NBF4 | *Decision making quality* | The system helps make decisions in completing tasks |

In brief, it can be seen that the indicators and measures were broken down from the model itself. In addition, the hypotheses were proposed based on the conceptual framework developed previously. Besides the adoption of the mobile learning context which may be the theoretical highlight of the model, the systematic formulation, its cohesive process, and the coherent stage of the model development may also be the methodological highlight of the study.

## The Instrument Testing

First, Table 2 elucidates profiles of the 50 respondents in the testing phase. The people dominant people were students who are the females (±70%), high school (±66%), public school (±78%), and the non-religious school (±72%) pupils. Meanwhile, Table 3 and Table 4 show results of the reliability and validity assessments of the indicators with 10 item rejections (i.e., INQ1, NBF1, SVQ1, SYQ1, SYQ2, SYQ5, SYU1, SYU2, USF1, and USF4) and the 21 reliable and valid indicators. The indicator reliability assessments were performed with cross loading (CL) threshold value 0.7, the internal consistency reliability with composite reliability (CR) threshold value 0.7, the convergent validity with average variance extracted (AVE) threshold value 0.5, and the discriminant validity with Fornell and Larcker's [38] square roots of AVEs.

TABLE 2 List of respondent profiles

| Profile | Name | f | % | Profile | Name | f | % |
|---|---|---|---|---|---|---|---|
| Gender | Male | 15 | 30 | Daily Frequency | 1 Time | 14 | 28 |
| | Female | 35 | 70 | | 2 Times | 12 | 24 |
| School Level | Elementary School | 4 | 8 | | 3 Times | 12 | 24 |
| | Middle School | 13 | 26 | | 4 Times | 4 | 8 |
| | High School | 33 | 66 | | >5 Times | 8 | 16 |
| School Type-1 | Public | 39 | 78 | Experience | <6 Months | 24 | 48 |
| | Private | 11 | 22 | Duration | 6-12 Months | 13 | 26 |
| School Type-2 | Religious School | 14 | 28 | | 1-2 Years | 8 | 16 |
| | Non-Religious School | 36 | 72 | | 2-3 Years | 2 | 4 |
| Regency | Jakarta | 16 | 32 | | >3 Years | 3 | 6 |
| | Depok | 4 | 8 | Content Type | Text | 7 | 14 |
| | Bogor | 7 | 14 | | Photo | 2 | 4 |
| | Tangerang | 13 | 26 | | Audio | 2 | 4 |
| | South Tangerang | 4 | 8 | | Video | 39 | 78 |
| | Bekasi | 6 | 12 | Success Level | Poor | 10 | 20 |
| Provider | Ruang Guru | 25 | 50 | | Fair | 1 | 2 |
| | Zenius | 21 | 42 | | Good | 19 | 38 |
| | Quipper | 4 | 8 | | Excellent | 20 | 40 |

TABLE 3. List of CLs, CRs, and AVEs

| Indicators | CLs | CRs | AVEs |
|---|---|---|---|
| INQ2 | 0.776 | 0.877 | 0.734 |
| INQ3 | 0.804 | | |
| INQ4 | 0.929 | | |
| INQ5 | 0.909 | | |
| NBF2 | 0.916 | 0.850 | 0.769 |
| NBF3 | 0.872 | | |
| NBF4 | 0.840 | | |
| SCT1 | 0.883 | 0.802 | 0.716 |
| SCT2 | 0.880 | | |
| SCT3 | 0.769 | | |
| SVQ2 | 0.889 | 0.605 | 0.714 |
| SVQ3 | 0.798 | | |
| SYQ3 | 0.928 | 0.816 | 0.845 |
| SYQ4 | 0.910 | | |
| SYU3 | 0.890 | 0.776 | 0.816 |
| SYU4 | 0.917 | | |
| UMT1 | 0.932 | 0.913 | 0.852 |
| UMT2 | 0.913 | | |
| UMT3 | 0.924 | | |
| USF2 | 0.940 | 0.868 | 0.884 |
| USF3 | 0.940 | | |

TABLE 4. The square roots of AVEs

| | INQ | NBF | SCT | SVQ | SYQ | SYU | UMT | USF |
|---|---|---|---|---|---|---|---|---|
| INQ | 0.857 | - | - | - | - | - | - | - |
| NBF | 0.549 | 0.877 | - | - | - | - | - | - |
| SCT | 0.673 | 0.486 | 0.846 | - | - | - | - | - |
| SVQ | 0.640 | 0.506 | 0.734 | 0.845 | - | - | - | - |
| SYQ | 0.679 | 0.420 | 0.725 | 0.666 | 0.919 | - | - | - |
| SYU | 0.660 | 0.677 | 0.608 | 0.576 | 0.621 | 0.903 | - | - |
| UMT | 0.647 | 0.798 | 0.538 | 0.542 | 0.583 | 0.783 | 0.923 | - |
| USF | 0.779 | 0.654 | 0.653 | 0.644 | 0.719 | 0.824 | 0.817 | 0.940 |

In summary, despite it was with 10 indicator rejections; results of the instrument testing expressed statistically the psychometric properties of the indicators [9-11]. Besides the interpretative evaluation, this statistical testing may also one of considerations for revising the questionnaires. As it was elucidated by Subiyakto, Rosalina [39], Carlsson, Ekstrand [18], and Liu, Li [40], in terms of a mixed questionnaires testing method.

## CONCLUSION

Considering to the research questions, two highlighted points of the study are regarding to the m-learning system success model and the reliability and validity its instruments. First, the model was developed based on the conceptual framework by adopting the IS success model and m-learning system context, combining above-mentioned model and theory, and adapting in terms of m-learning system success model. Second, 10 of the proposed indicators were rejected in the statistical assessments. It can be clearly seen that despite the rejections; results of the instrument testing expressed statistically the psychometric properties of the indicators. Furthermore, besides the adoption of the mobile learning context which may be the theoretical highlight of the model, the systematic formulation, its cohesive process, and the coherent stage of the model development may also be the methodological highlight of the study. Of course, the uses of the data, methodological points, and findings of the study cannot be generalized for the other studies. Therefore, it may be one of consideration for the other ones. Practically, besides the interpretative evaluation, this statistical testing may also one of considerations for revising the questionnaires.

# REFERENCES

1. H. Crompton and D. Burke, *Computers & Education*, **123**, pp. 53-64 (2018).
2. S.J.H. Chester, J.H.Y. Stephen, H.C.C. Tosti and Y.S.S. Addison, *Journal of Educational Technology & Society*, **19** (1), pp. 263-276 (2016).
3. J. Keengwe and M. Bhargava, *Education and Information Technologies*, **19** (4), pp. 737-46 (2014).
4. D. Sulisworo and M. Toifur, *International Journal of Mobile Learning and Organisation*, **10** (3), pp. 159-170 (2016).
5. V.A. Anfara Jr and N.T. Mertz, *Theoretical Frameworks in Qualitative Research* (Sage publications, 2014).
6. A. Subiyakto, *Bulletin of Electrical Engineering and Informatics*, **7** (3), pp. 400-410 (2018).
7. F. Abdullah and R. Ward, *Computers in Human Behavior*, **56**, pp. 238-256 (2016).
8. A. Subiyakto, M.R. Juliansyah, M.C. and A. Susanto, in *2018 6th International Conference on Cyber and IT Service Management (CITSM)* (IEEE, 2018), pp. 1-5.
9. J. Hair, *Industrial Management & Data Systems*, **117** (3), pp. 442-458 (2017).
10. M. Sarstedt, C.M. Ringle and J.F. Hair, *Long Range Planning*, **47** (3), pp. 132-137 (2014).
11. N. Urbach and F. Ahlemann, *JITTA: Journal of Information Technology Theory and Application*, **11** (2), pp. 5-40 (2010).
12. D.M. Eddy, W. Hollingworth, J.J. Caro, J. Tsevat, K.M. McDonald and J.B. Wong, *Medical Decision Making*, **32** (5), pp. 733-743 (2012).
13. A. Subiyakto, A.R. Ahlan, S.J. Putra and M. Kartiwi, *SAGE Open*, **5** (2), pp. 1-14 (2015).
14. S. Petter, W. DeLone and E. McLean, European Journal of Information Systems, **17** (3), pp. 236-263 (2008).
15. H.B. Seta, T. Wati, A. Muliawati and A.N. Hidayanto, *Indonesian Journal of Electrical Engineering and Informatics (IJEEI)*, **6** (3), pp. 281-291 (2018).
16. K.D. Chen and P.K. Chen, *Asia Pacific Education Review*, **18** (4), pp. 439-449 (2017).
17. W.C. Wu and Y.H. Perng, *Eurasia Journal of Mathematics, Science and Technology Education*, **12** (6), (2016).
18. I. Carlsson, E. Ekstrand, M. Åström, K. Stihl and M. Arner, Content validity, construct validity and magnitude of change for the eight-item HAKIR questionnaire - a patient reported outcome in the Swedish national healthcare quality registry for hand surgery, Research Square, 2019. Available from: https://www.researchsquare.com/article/rs-8743/v1.
19. M. Glaser, D. Lengyel, C. Toulouse and S. Schwan, *Educational Technology Research and Development*, **65** (5), pp. 1135-1151 (2017).
20. T. Jagušt and I. Botički, *Journal of Computers in Education*, **6** (3), pp. 335-362 (2019).
21. K.Y Kwahk, H. Ahn and Y.U. Ryu, *International Journal of Information Management*, **38** (1), pp. 64-76 (2018).
22. M. Martinsuo, *International Journal of Project Management*, **31** (6), pp. 794-803 (2013).
23. O. Zwikael and M. Ahn, *Risk analysis*, **31** (1), 25-37 (2011).
24. W.H. DeLone and E.R. McLean, *Journal of Management Information Systems*, **19** (4), pp. 9-30 (2003).
25. N. Urbach and B. Müller, *The updated DeLone and McLean Model of Information Systems success* (Springer, New York, 2012), pp. 1-18.
26. U. Haryaka, F. Agus and A.H. Kridalaksana, *Procedia Computer Science*, **116**, pp. 373-380 (2017).
27. B.K. Riasti and A. Nugroho, *Journal of Physics: Conference Series*, **1339**, p. 012063 (2019).
28. A. Subiyakto, N.A. Hidayah, G. Gusti and M.A. Hikami, *Information*, **10** (2), p. 79 (2019).
29. D.I. Sensuse and D.B. Napitupulu, *Indonesian Journal of Electrical Engineering and Computer Science*, **7** (2), pp. 466-473 (2017).
30. L.D. Krisnawati, *New Horizons in Education*, **57** (3), pp. 74-81 (2009).
31. S.K. Alzu'B and S. Hassan, *Asian Journal of Information Technology*, **15** (1), pp. 113-121 (2016).

32. W.H. Wu, Y.C. Jim Wu, C.Y. Chen, H.Y. Kao, C.H. Lin and S.H. Huang, *Computers & Education*, **59** (2), pp. 817-827 (2012).
33. M.A. Virtanen, E. Haavisto, E. Liikanen and M. Kääriäinen, *Education and Information Technologies*, **23** (2), pp. 985-998 (2018).
34. W. Bhuasiri, O. Xaymoungkhoun, H. Zo, J.J. Rho and A.P. Ciganek, *Computers & Education*, **58** (2), pp. 843-855 (2012).
35. A.E.E. Sobaih, M.A. Moustafa, P. Ghandforoush and M. Khan, *Computers in Human Behavior*, **58**, pp. 296-305 (2016).
36. C. Snelson, *Learning, Media and Technology*, **43** (3), pp. 294-306 (2018).
37. S.H. Jeong, H. Kim, J.Y. Yum and Y. Hwang, *Computers in Human Behavior*, **54**, 10-17 (2016).
38. C. Fornell and D.F. Larcker, *Journal of Marketing Research*, pp. 382-388 (1981).
39. A. Subiyakto, R. Rosalina, M.C. Utami, N. Kumaladewi and S.J. Putra, in *5th International Conference on Information Technology for Cyber and IT Service Management (CITSM)* (IEEE, 2017).
40. P. Liu, Q. Li, J. Bian, L. Song and X. Xiahou, *International Journal of Environmental Research and Public Health*, **15** (7), P. 1359 (2018).